\documentclass{article}

\usepackage{hyperref}

\usepackage{tikz} 

\usepackage{epsfig}
\usepackage{comment}
\usepackage{booktabs}

\usepackage{amsmath, amssymb}
\usepackage{nicefrac}
\usepackage{subcaption}

\usepackage{ntheorem}

\usepackage{tikz}
\usetikzlibrary{patterns}
\usetikzlibrary{arrows.meta}
\usetikzlibrary{calc, math}
\usepackage{gnuplot-lua-tikz}
\usepackage{placeins}
\usepackage{float}

\usepackage[shortlabels]{enumitem}
\usepackage{multirow}

\makeatletter
\g@addto@macro\@floatboxreset\centering
\makeatother

\renewcommand{\vec}[1]{\mathbf{#1}}
\newcommand{\mat}[1]{\mathbf{#1}}
\newcommand{\relu}{\operatorname{ReLU}}
\newcommand{\E}{\operatorname{E}}
\newcommand{\prob}{\operatorname{P}}
\newcommand{\erf}{\operatorname{erf}}
\newcommand{\diff}{\mathrm{d}}
\usepackage[accepted]{icml2020}
\icmltitlerunning{On the security relevance of weights in deep learning}

\begin{document}
\twocolumn[
\icmltitle{On the security relevance of weights in deep learning}


\icmlsetsymbol{equal}{*}

\begin{icmlauthorlist}
\icmlauthor{Kathrin Grosse\textsuperscript{*}}{cis,sic}
\icmlauthor{Thomas A. Trost\textsuperscript{*}}{lsv,sic}
\icmlauthor{Marius Mosbach}{lsv,sic}
\icmlauthor{Michael Backes}{cis}
\icmlauthor{Dietrich Klakow}{lsv,sic}
\end{icmlauthorlist}

\icmlaffiliation{cis}{CISPA Helmholtz Center for information Security}
\icmlaffiliation{sic}{Saarland University, Saarland Informatics Campus}
\icmlaffiliation{lsv}{Spoken Language Systems (LSV)}

\icmlcorrespondingauthor{Kathrin Grosse}{kathrin.grosse@cispa.saarland}

\icmlkeywords{Adversarial Machine Learning, training time attacks}

\vskip 0.3in
]

\printAffiliationsAndNotice{\icmlEqualContribution}


\begin{abstract}
  Recently, a weight-based attack on stochastic gradient descent inducing overfitting has been proposed. 
  We show that the threat is broader: A task-independent permutation on the initial weights suffices to limit the achieved accuracy
  to for example 50\% on the Fashion MNIST dataset from initially more than $90$\%. 
These findings are confirmed on MNIST and CIFAR.
We formally confirm that the attack succeeds with high likelihood and does not depend on the data.
Empirically, weight statistics and loss appear unsuspicious, making it hard to
 detect the attack if the user is not aware. 
  Our paper is thus a call for action to acknowledge the importance of the initial weights in deep learning.
\end{abstract}


\section{Introduction}



Deep learning is ubiquitous. Applications range from computer
vision~\cite{he-etal:2015}, autonomous driving, natural language
processing~\cite{manning:2015} and Malware
detection~\cite{DBLP:conf/malware/SaxeB15} to
health-care~\cite{miotto2017deep}. Many applications are
fueled by large amounts of collected data, and the performance
is impressive.

However, security concerns about such a wide application
of machine learning (ML) are raised \cite{biggio:2018}. One example
among many is the threat of \emph{poisoning}. The corresponding attacker
manipulates the training data to alter the resulting
classifier's accuracy 
\cite{rubinstein2009antidote,mei2015using,biggio2012poisoning,liu2017trojaning,shafahi2018poison} by targeting the training of the classifier.
Recent work has tailored poisoning to deep neural networks~\cite{liu2017trojaning,zhu2019transferable,liu2019bad}.
Due to their flexibility, these models are harder to target than for example an SVM. Our attack is loosely related to these approaches.

Training and in particular initialization of deep neural
networks is still based on heuristics, such as breaking
symmetries in the network, and avoiding that gradients vanish or
explode \cite{bengio:1994,pascanu-mikolov-bengio:2013}.
State of the art approaches rely on the idea that given a random
initialization, the variance of weights is particularly important
\cite{hanin:2018, hanin-rolnick:2018} and determines the dynamics of
the networks \cite{kadmon-sompolinsky:2015, poole-etal:2016}. In
accordance with this, weights are nowadays usually simply drawn from
some zero-centered (and maybe cut-off) Gaussian distribution with
appropriate variance \cite{giryes-etal:2016}, while the biases are
often set to a constant. The order of the weights is typically not considered, so an adversarial (or simply unlucky) permutation with particularly bad properties has a good chance of being overseen, if the user is not aware of this kind of problem. 

\textbf{Contributions.}
We propose a training-data-independent attack on the performance of neural networks that underlines the importance of the initial weights.
Specifically, we show ways to permute initial weights before training (such that all statistics are preserved and seem inconspicuous) that effectively reduce the network capacity, implying decreased accuracy and increased training
time. More concretely, on the MNIST benchmark, where benign accuracy
is easily $>98$\%, the attacker is able to limit the accuracy to
$50$\%. On Fashion MNIST, she reduces the accuracy from $>90$\% to
slightly more than $50$\%. For CIFAR, the accuracy of our simple LeNet~\cite{lecun1998gradient} model is analogously
reduced from $65$\% to $50$\%.
Our formal analysis confirms the attack's independence of the training data.



%

\subsection{Related Work}

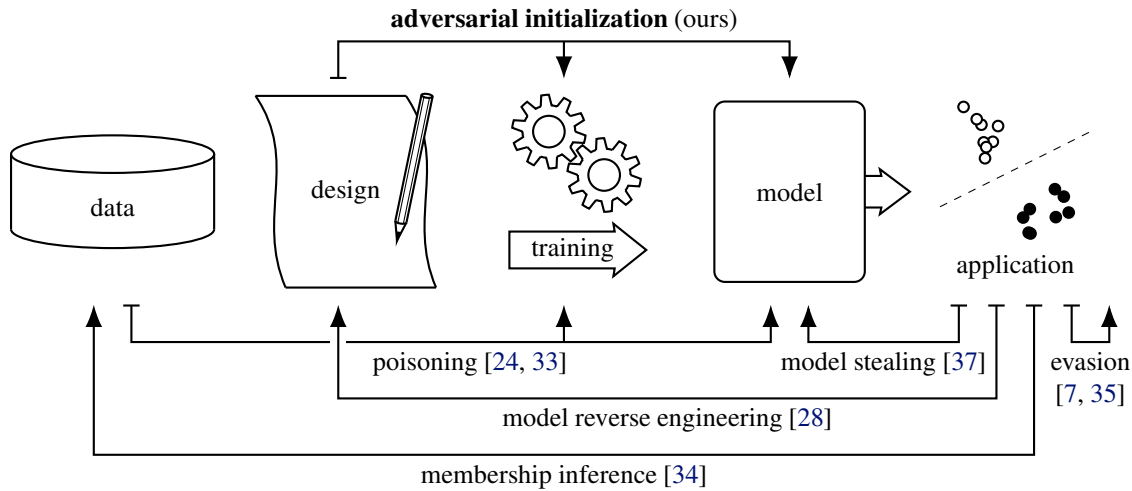
\begin{figure*}[t]
\centering
\begin{tikzpicture}
  \tikzset{myline/.style = {thick, black}}
  \tikzset {
    partial ellipse/.style args={#1:#2:#3}{
      insert path={+ (#1:#3) arc (#1:#2:#3)}
    }
  }

  \begin{scope}[xshift=-6cm,xscale=0.9]
    \draw [myline] (0, 0.5) ellipse (1.5cm and 0.25cm);
    \draw [myline] (0, -0.5) [partial ellipse=180:360:1.5cm and 0.25cm];
    \draw [myline] (-1.5, -0.5) -- (-1.5, 0.5);
    \draw [myline] (1.5, -0.5) -- (1.5, 0.5);
    \draw node at (0, -0.2) {data};    
  \end{scope}

  \begin{scope}[xshift=-3.1cm,yscale=0.85]
    \draw [myline] (-1,1.5) -- (1,1.5) edge[out=-60, in=110] (1.4,-1.5);
    \draw [myline] (1.4,-1.5) -- (-0.6,-1.5) edge[out=110, in=-60] (-1,1.5);
    \draw node at (0.2, 0.) {design};
    \begin{scope}[rotate=-10,xshift=1cm,yshift=0.7cm]
      \fill [fill=white] (-0.075,-1) rectangle (0.075,1);
      \draw [myline] (0.075,-1) -- (0.075,1);
      \draw [myline] (-0.075,-1) -- (-0.075,1);
      \draw [myline] (0, 1) ellipse (0.075cm and 0.03cm);
      \draw [myline, fill=white] (-0.075,-1) -- (0,-1.3) -- (0.075,-1);
      \draw [myline] (0, -1) [partial ellipse=180:360:0.075cm and 0.03cm];
      \draw [] (0.0375,-1.025) -- (0.0375,0.99);
      \draw [] (-0.0375,-1.025) -- (-0.0375,0.99);
      \draw [myline, fill=black] (-0.01875,-1.175) -- (0,-1.3) -- (0.01875,-1.175) -- cycle;
    \end{scope}
  \end{scope}

  \begin{scope}[xshift=-0.2cm,yshift=0.8cm,scale=0.7]
    \draw[myline]
    \foreach \i in {1,2,...,10} {%
      [rotate=(\i-1)*36]  (0:0.5)  arc (0:12:0.5) -- (18:0.7)  arc (18:30:0.7) --  (36:0.5)
    };
    \draw [myline] (0,0) circle (0.3);
    \begin{scope}[xshift=1.05cm,yshift=-0.825cm]
      \draw[myline]
      \foreach \i in {1,2,...,10} {%
        [rotate=(\i-1)*36]  (0:0.5)  arc (0:12:0.5) -- (18:0.7)  arc (18:30:0.7) --  (36:0.5)
      };
      \draw [myline] (0,0) circle (0.3);
    \end{scope}
    \begin{scope}[xshift=0.25cm, yshift=-2.25cm]
      \draw [myline] (-1.,0.35) -- (1,0.35) -- (0.975,0.5) -- (1.6, 0) -- (0.975,-0.5) -- (1,-0.35) -- (-1,-0.35) -- cycle;
      \draw node at (0.2,0.) {training};
    \end{scope}
  \end{scope}

  \begin{scope}[xshift=3cm]
    \draw [myline, rounded corners] (-1,-1.2) rectangle (1,1.2);
    \draw node at (0,0) {model};
    \draw [myline] (1., 0.21) -- (1.22, 0.21) -- (1.2, 0.3) -- (1.6, 0) -- (1.2, -0.3) -- (1.22, -0.21) -- (1., -0.21);
  \end{scope}

  \begin{scope}[xshift=6cm,yshift=0.3cm]
    \foreach \x in {1, ..., 8}{
      \draw [myline,fill=white] (0.4*rand-0.5,0.4*rand+0.5) circle (0.07);
      \draw [myline,fill=black] (0.4*rand+0.5,0.4*rand-0.5) circle (0.07);
    }
    \draw [dashed] (-1,-0.5) -- (1,0.5);
    \draw node at (0,-1.3) {application};
  \end{scope}

  \newcommand{\attackarrow}[4]{
    \draw [thick, |-{Latex[length=3mm]}] (#1, -1.5) -- (#1, -1.5 - #3) -- node[below,align=center]{#4} (#2, -1.5 - #3) -- (#2, -1.5);
  }
  \newcommand{\attackarrowa}[4]{
    \draw [thick, |-{Latex[length=3mm]}] (#1, 1.5) -- (#1, 1.5 + #3) -- node[above]{#4} (#2, 1.5 + #3) -- (#2, 1.5);
  }

  \attackarrow{-5.75}{2.75}{0.5}{\hspace{5mm}poisoning~\cite{liu2017trojaning,shafahi2018poison}}
  \fill [fill=white] (-3.1, -1.8) rectangle (-2.9, -2.2);
  \attackarrow{5.25}{3.25}{0.5}{model stealing \cite{DBLP:conf/uss/TramerZJRR16}}
  \attackarrow{5.75}{-3.}{1.25}{model reverse engineering \cite{joon2018towards}}
  \attackarrow{6.25}{-6.25}{2.00}{membership inference \cite{2016arXiv161005820S}}
  \attackarrow{6.75}{7.25}{0.5}{evasion\\\cite{Dalvi:2004:AC:1014052.1014066,szegedy2013intriguing}}
  \attackarrowa{-3}{3}{0.5}{\textbf{adversarial initialization} (ours)}
  \draw [thick, -{Latex[length=3mm]}] (0,2) -- (0,1.5);
  \draw [thick, -{Latex[length=3mm]}] (0,-2) -- (0,-1.5);

\end{tikzpicture}
\caption{An overview of attacks on Machine Learning.}
\label{fig:pipeline}
\end{figure*}

We give an overview over attacks on ML in Fig.~\ref{fig:pipeline}.
To the best of our knowledge, there is little work targeting the initial
weights before training.
Orthogonal works explore poisoning for deep learning.
Due to the flexibility of the models, these attacks are currently
limited to the misclassification of individual points~\cite{zhu2019transferable} or the introduction
of backdoors~\cite{liu2017trojaning,shafahi2018poison,tan2019bypassing}.
Such a backdoor pattern is small, yet tricks the model
into reliably outputting an attacker-chosen class. These attacks rely on
altering the training data. 
Also orthogonally, Cheney et al.~\yrcite{cheney2017robustness} investigate adversarial weight perturbations at test time (not at training time of the initial weights).
In general, benign hardware failures during training have been studied as well~\cite{2017arXiv170405396V}.

Closest to our work, Liu et al.~\yrcite{liu2019bad} extended training time-attacks to the
 weights of an SGD-trained model which consecutively over-fits the data.
There are several differences to our contribution:
(1) our attacks are independent of
the optimizer and other hyper-parameters, and (2) the damage of decreased accuracy is more severe than overfitting. Furthermore, (3) our
attack is also more stealthy, as the statistics of the original weights are preserved, and (4) our attacks take place \emph{before} training.


\section{Adversarial Initialization}
In this work, we introduce attacks that alter the initial weights of a neural network.
The goal of the attacker is to decrease accuracy drastically, or to increase training time.
Ideally, this is done in a stealthy way: if the victim spots the attack, no harm is done.

Before we discuss specifics and the generalization of our attacks, we
motivate our approach by discussing its most basic version. The following equation represents two consecutive layers in a
fully connected feed-forward network with weight matrices
$\mat A \in \mathbb{R}^{m\times n}$ and
$\mat B \in \mathbb{R}^{\ell\times m}$, corresponding biases $\vec a \in \mathbb{R}^{m}$
as well as $\vec b \in \mathbb{R}^{\ell}$, and ReLU activation functions:
\begin{equation}\label{eq:consecutive-layers}
  \vec y = \relu\big(\mat B\relu(\mat A\vec x + \vec a) + \vec b\big)
\end{equation}
This vulnerable structure or similar vulnerable structures (like two consecutive
convolutional layers) 
can be
found in a plethora of typical DNN architectures. We assume that the
neurons are represented as column vectors. The formulation for
  a row vector is completely analogous.

We further assume that the components of $\vec x$ are
positive. This corresponds to the standard normalization of the input data
between $0$ and $1$. For input vectors $\vec x$ resulting from
the application of previous layers it is often reasonable to expect an
approximately normal distribution with the same characteristics for
all components of $\vec x$. This assumption is (particularly) valid
for wide previous layers with randomly distributed weights because the
sum of many independent random variables is an approximately normally
distributed random variable due to the central limit theorem
\cite{poole-etal:2016}.

The idea behind our approach is to make many components of $\vec y$
vanish with high probability and is best illustrated by means of the sketches in equation~\ref{eq:ax} and
equation~\ref{eq:bx}. The components of the matrices and vectors are depicted
as little squares. Darker colors mean larger values. In addition,
hatched squares indicate components with a high probability of being
zero.

In matrix $\mat A$, the largest components of the original matrix are
all randomly distributed in the lower $(1 - r_A)m$ rows. $r_A\in\{\frac 1 m, \frac 2 m, ..., 1\}$ controls the fraction of rows that are filled with the ``small'' values. The small and often negative components are randomly distributed in the upper $r_Am$
rows. The products of these negative rows with the positive $\vec x$ are
likely negative.  If the
bias $\vec a$ is not too large, the resulting vector will have many
zeros in the upper rows due to the ReLU-cutoff.
\begin{equation}\label{eq:ax}
  \relu
  \left(
    \underbrace{
      \begin{bmatrix}
        \begin{tikzpicture}[scale=0.8]
          \foreach \y in {0, 1} {
            \foreach \x in {0, ..., 3} {
              \pgfmathparse{0.3*(rnd-0.5) + 0.2}
              \definecolor{darkrnd}{rgb}{\pgfmathresult,\pgfmathresult,\pgfmathresult}
              \fill[darkrnd] (0.5*\x, 0.5*\y) rectangle (0.5*\x + 0.4, 0.5*\y+0.4);
            }
          }
          \foreach \y in {2, 3, 4} {
            \foreach \x in {0, ..., 3} {
              \pgfmathparse{0.2*(rnd-0.5) + 0.85}
              \definecolor{lightrnd}{rgb}{\pgfmathresult,\pgfmathresult,\pgfmathresult}
              \fill[lightrnd] (0.5*\x, 0.5*\y) rectangle (0.5*\x + 0.4, 0.5*\y+0.4);
            }
          }
          \fill[white] (0, 2.5) rectangle (0.4, 2.55); 
        \end{tikzpicture}
      \end{bmatrix}}_{\text{matrix}\; \mat A}
    \underbrace{
      \begin{bmatrix}
        \begin{tikzpicture}[scale=0.8]
          \foreach \y in {0, ..., 3} {
            \pgfmathparse{0.2*(rnd-0.5) + 0.5}
            \definecolor{midrnd}{rgb}{\pgfmathresult,\pgfmathresult,\pgfmathresult}
            \fill[midrnd] (0, 0.5*\y) rectangle (0.4, 0.5*\y+0.4);
          }
          \fill[white] (0, 2.) rectangle (0.4, 2.05); 
        \end{tikzpicture}        
      \end{bmatrix}
    }_{\vec x} + \vec a
  \right)
  =
  \begin{bmatrix}
    \begin{tikzpicture}[scale=0.8]
      \foreach \y in {0, 1} {
        \foreach \x in {0} {
          \fill[black!90] (0.5*\x, 0.5*\y) rectangle (0.5*\x + 0.4, 0.5*\y+0.4);
        }
      }
      \foreach \y in {2, 3, 4} {
        \foreach \x in {0} {
          \fill[pattern=north west lines] (0.5*\x, 0.5*\y) rectangle (0.5*\x + 0.4, 0.5*\y+0.4);
        }
      }
      \fill[white] (0, 2.5) rectangle (0.4, 2.55); 
    \end{tikzpicture}
  \end{bmatrix}
\end{equation}

Next, a similar approach can be used with matrix $\vec B$ to eliminate
the remaining positive components. Let $r_B$ control the fraction of
``small'' columns of $\mat B$.
\begin{equation}\label{eq:bx}
  \relu
\left(
  \underbrace{
    \begin{bmatrix}
      \begin{tikzpicture}[scale=0.8]
        \foreach \y in {0, ..., 3} {
          \foreach \x in {0, 1} {
            \pgfmathparse{0.3*(rnd-0.5) + 0.2}
            \definecolor{darkrnd}{rgb}{\pgfmathresult,\pgfmathresult,\pgfmathresult}
            \fill[darkrnd] (0.5*\x, 0.5*\y) rectangle (0.5*\x + 0.4, 0.5*\y+0.4);
          }
        }
        \foreach \y in {0, ..., 3} {
          \foreach \x in {2, 3, 4} {
            \pgfmathparse{0.2*(rnd-0.5) + 0.85}
            \definecolor{lightrnd}{rgb}{\pgfmathresult,\pgfmathresult,\pgfmathresult}
            \fill[lightrnd] (0.5*\x, 0.5*\y) rectangle (0.5*\x + 0.4, 0.5*\y+0.4);
          }
        }
        \fill[white] (0, 2.) rectangle (0.4, 2.05); 
      \end{tikzpicture}
    \end{bmatrix}
  }_{\text{matrix}\; \mat B}
  \begin{bmatrix}
    \begin{tikzpicture}[scale=0.8]
      \foreach \y in {0, 1} {
        \foreach \x in {0} {
          \fill[black!90] (0.5*\x, 0.5*\y) rectangle (0.5*\x + 0.4, 0.5*\y+0.4);
        }
      }
      \foreach \y in {2, 3, 4} {
        \foreach \x in {0} {
          \fill[pattern=north west lines] (0.5*\x, 0.5*\y) rectangle (0.5*\x + 0.4, 0.5*\y+0.4);
        }
      }
      \fill[white] (0, 2.5) rectangle (0.4, 2.55); 
    \end{tikzpicture}
  \end{bmatrix} + \vec b
\right)
=
\underbrace{
  \begin{bmatrix}
    \begin{tikzpicture}[scale=0.8]
      \foreach \y in {0, ..., 3} {
        \foreach \x in {0} {
          \fill[pattern=north west lines] (0.5*\x, 0.5*\y) rectangle (0.5*\x + 0.4, 0.5*\y+0.4);
        }
      }
      \fill[white] (0, 2.) rectangle (0.4, 2.05); 
    \end{tikzpicture}
  \end{bmatrix}
}_{\vec y}
\end{equation}

In summary, we concentrate the positive contributions in a few places
and ``cross'' $\vec A$ and $\vec B$ in order to annihilate them.  For
the typical case of weights drawn from a zero mean
distribution, $r_A = r_B = \frac 1 2$ effectively kills all the
neurons and makes training impossible.

The probability for obtaining a matrix like $\mat A$ in equation~\ref{eq:ax}
by chance is very small and given by $\nicefrac{\left((r_Amn)!\left((1-r_A)mn\right)!\right)}{(mn)!}$.

\subsection{Detailed Description of Attacks}
With the general idea of our attack in mind, we can now discuss
specifics. A complete blockade of the entire network obviously
contradicts the idea of stealthiness because at least some learning is
expected by the user. The prototypical attack must thus be
``weakened'' in a controlled manner to comply with the specification
of an attack. Towards this end we introduce specific implementations
of the idea that can actually be used in practice. Finally, we discuss
the special case of convolutional layers. 

\textbf{Soft Knockout Attack.} 
The first way of controlling the network capacity is by varying $r_A$
and $r_B$ in such a way that some but not all of the neurons have some
non-vanishing probability of being non-zero. This is achieved by
choosing $r_A < \nicefrac 1 2$ or $r_B < \nicefrac 1 2$, respectively
$r_A \gg \nicefrac 1 2$ or $r_B \gg \nicefrac 1 2$.

\textbf{Shift Attack.} Alternatively, we can
choose $r_A = r_B = \nicefrac 1 2$ and shift the columns of $\mat B$
periodically by $s$ positions. In a fully connected network, this
corresponds to $s$ active neurons, yielding good control
over the final capacity of the network.

\textbf{Algorithmic Procedure.} We formalize both approaches in Algorithm~\ref{alg::bleeding} for fully connected
layers. 
 In both cases, the attacker only alters one weight matrix at a time, in the
order one would initialize the weights in the model.
To perturb a weight matrix, we first obtain the $(100 \times r)$\%
smallest weights, denoted as $\mat S$ (line 4
). We
denote the remaining, larger weights as $\mat L$ (line 5
).
For the shift attack, we instead split the weight matrix in
negative (line 7
) and non-negative (line 8
) components. 
Depending on the variable
\emph{cross} which is flipped each round (line 18
), we reorder the weights. 
For the first and all
unevenly indexed matrices, we align the small weights in the upper
rows and then fill up with the larger weights (line 13
). For all evenly indexed matrices, we cross
the components: The first columns contain the large weights, and we
fill up with the small weights (line 11
). 
When using the shift attack, In all uneven cases, we additionally shift the rows by $s$ positions to
activate $s$ neurons (line 15
).
In this
formalization, we skip the details of the exact reshaping operations
needed to obtain matrices of the correct shape. We also leave aside that
columns and rows might be filled partially with small and large
weights.

\begin{algorithm}
  \caption{\textbf{Soft Knockout/Shift Attack.} Given a stream of weights
    $\mathcal{W} = \{\mat{W}_1, ...\}$, parameter $r \in[0,1]$, and $s \in \mathbb{N}$,
    outputs permuted, harmful weights impeding training.}\label{alg::bleeding}
  \begin{algorithmic}[1]
    \REQUIRE $\mathcal W$, $r$, $s$
    \STATE cross $\leftarrow$ False
    \FOR{$\mat{W}_i \in \mathcal W$}
    \IF{soft knockout attack}
    \STATE $\mat S \leftarrow$ smallest $r|\mat{W}_i|$ components of $\mat{W}_i$\label{algo:blSortS}
    \STATE $\mat L \leftarrow$ largest $(1 - r)|\mat{W}_i|$ components of $\mat{W}_i$ \label{algo:blSortL}
    \ELSE[\emph{prepare shift attack}]
    \STATE $\mat S \leftarrow$ negative components of $\mat{W}_i$ \label{algo:SSortS}
    \STATE $\mat L \leftarrow$ non-negative components of $\mat{W}_i$\label{algo:SSortL}
    \ENDIF
    \IF{cross}
    \STATE $\mat{W}_i \leftarrow \begin{pmatrix} \mat L & \mat S\end{pmatrix}$\label{algo:blCrossCase}
    \ELSE
    \STATE $\mat{W}_i \leftarrow \begin{pmatrix} \mat S \\ \mat L\end{pmatrix}$\label{algo:blNonCrossCase}
    \IF{shift attack}
    \STATE shift rows of $\mat{W}_i$ by $s$ positions periodically\label{algo:Sshift}
    \ENDIF
    \ENDIF
    \STATE cross $\leftarrow \neg $ cross \label{algo:blswapCross}
    \ENDFOR
  \end{algorithmic}
\end{algorithm}

\begin{figure}
  \centering
  \begin{tikzpicture}
  \tikzset{myarrow/.style={-{Latex}, thick}}
  \tikzset{myhiddennode/.style={}}
  \def\mylightcl{0.2*(rnd-0.5) + 0.85}
  \def\mydarkcl{0.3*(rnd-0.5) + 0.2}
  \begin{scope}[yshift=-0.4cm]
    \foreach \y in {0, ..., 3} {
      \foreach \x in {0, ..., 3} {
        \pgfmathparse{0.2*(rnd-0.5) + 0.5}
        \definecolor{midrnd}{rgb}{\pgfmathresult,\pgfmathresult,\pgfmathresult}
        \fill[midrnd] (0.2*\x, 0.2*\y) rectangle (0.2*\x + 0.2, 0.2*\y+0.2);
      }
    }
    \draw [step=0.2] (0, 0) grid (0.8, 0.8);
    \draw [thick] (0, 0) rectangle (0.8, 0.8);
    \node [above, align=center] at (0.4, .8) {input};
    \node[myhiddennode] (input) at (0.8, 0.4) {};
  \end{scope}

  \begin{scope}[xshift=1.6cm,yshift=-1.8cm]
    \foreach \y in {0, ..., 3} {
      \begin{scope}[yshift=\y cm]
        \foreach \iy in {0, ..., 2} {
          \foreach \ix in {0, ..., 2} {
            \ifnum \y>1
            \pgfmathparse{\mylightcl}
            \else
            \pgfmathparse{\mydarkcl}
            \fi
            \definecolor{mycl}{rgb}{\pgfmathresult,\pgfmathresult,\pgfmathresult}
            \fill[mycl] (0.2*\ix, 0.2*\iy) rectangle (0.2*\ix + 0.2, 0.2*\iy+0.2);
          }
        }
        \draw [step=0.2] (0, 0) grid (0.6, 0.6);
        \draw [thick] (0, 0) rectangle (0.6, 0.6);
        \node[myhiddennode] (f1\y in) at (0, 0.3) {};
        \draw [myarrow] (input) -- (f1\y in);
        \node[myhiddennode] (f1\y out) at (0.6, 0.3) {};
      \end{scope}
    }
    \node [above=0.3em] at (0.3, 3.6) {4 filters};
  \end{scope}

  \begin{scope}[xshift=3.0cm,yshift=-1.9cm]
    \foreach \y in {0, ..., 3} {
      \begin{scope}[yshift=\y cm]
        \ifnum \y>1
        \fill [pattern=north east lines] (0, 0) rectangle (0.8, 0.8);
        \else
        \foreach \iy in {0, ..., 3} {
          \foreach \ix in {0, ..., 3} {
            \pgfmathparse{\mydarkcl}
            \definecolor{mycl}{rgb}{\pgfmathresult,\pgfmathresult,\pgfmathresult}
            \fill[mycl] (0.2*\ix, 0.2*\iy) rectangle (0.2*\ix + 0.2, 0.2*\iy+0.2);
          }
        }
        \fi
        \draw [step=0.2] (0, 0) grid (0.8, 0.8);
        \draw [thick] (0, 0) rectangle (0.8, 0.8);
        \node[myhiddennode] (c1\y in) at (0, 0.4) {};
        \draw [myarrow] (f1\y out) -- (c1\y in);
        \node[myhiddennode] (c1\y out) at (0.8, 0.4) {};
      \end{scope}
    }
    \node [above, align=center] at (0.4, 3.8) {intermediate\\layer};
  \end{scope}

  \begin{scope}[xshift=4.7cm,yshift=-1.8cm]
    \foreach \y in {0, ..., 3} {
      \foreach \x in {0, ..., 3} {
        \begin{scope}[xshift=\x cm, yshift=\y cm]
          \foreach \iy in {0, ..., 2} {
            \foreach \ix in {0, ..., 2} {
              \ifnum \y>1
              \pgfmathparse{\mydarkcl}
              \else
              \pgfmathparse{\mylightcl}
              \fi
              \definecolor{mycl}{rgb}{\pgfmathresult,\pgfmathresult,\pgfmathresult}
              \fill[mycl] (0.2*\ix, 0.2*\iy) rectangle (0.2*\ix + 0.2, 0.2*\iy+0.2);
            }
          }
          \draw [step=0.2] (0, 0) grid (0.6, 0.6);
          \draw [thick] (0, 0) rectangle (0.6, 0.6);
        \end{scope}
      }
      \node[myhiddennode] (f2\y in) at (-0.1, 0.3+\y) {};
      \draw [myarrow] (c1\y out) -- (f2\y in);
    }
    \node [above=0.5em] at (1.8, 3.6) {4 filters with 4 channels};
    \foreach \x in {0, ..., 3} {
      \node[myhiddennode] (f2\x out) at (\x+0.3, -0.1) {};
    }
    \foreach \y in {0, ..., 2} {
      \foreach \x in {0, ..., 3} {
        \node[] at (0.3+\x, 0.8+\y) {+};
      }
    }
    \foreach \x in {0, ..., 3} {
      \draw (\x-0.1, -0.1) rectangle (\x+0.7, 3.7);
    }
  \end{scope}

  \begin{scope}[xshift=4.6cm,yshift=-3.5cm]
    \foreach \x in {0, ..., 3} {
      \begin{scope}[xshift=\x cm]
        \draw [thick, pattern=north east lines] (0, 0) rectangle (0.8, 0.8);
        \draw [step=0.2] (0, 0) grid (0.8, 0.8);
        \node[myhiddennode] (c2\x in) at (0.4, 0.8) {};
        \draw [myarrow] (f2\x out) -- (c2\x in);
      \end{scope}
    }
    \node [left=0.5em] at (0.0, 0.4) {output};
  \end{scope}

\end{tikzpicture}
  \caption{Sketch of our attack on convolutional layers. Light colors
  indicate small (negative) values, dark gray larger (positive) values.
  Hatched fields indicate that this output is likely to be $0$ if ReLU
  is used as an activation.} 
  \label{fig:conv-attack}
\end{figure}
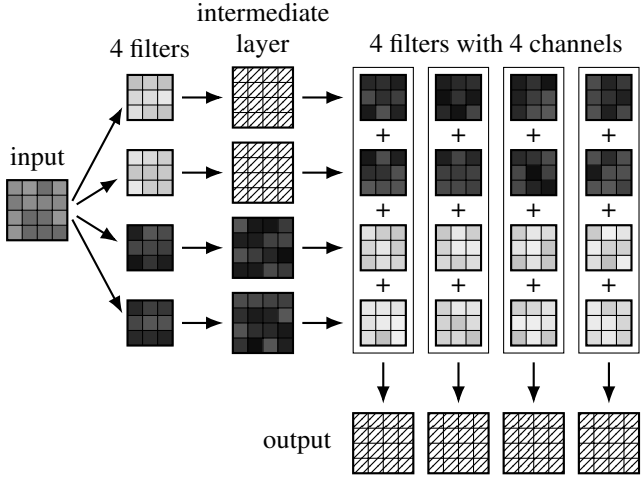

\textbf{Adversarial Initialization for Convolutions.} 
Particular care has to be taken when attacking convolutional
layers. Yet, the idea of weight permutation and matrix crossing
works in a very similar way. We formalize the attacks for convolutional weights
represented as
$4$-dimensional tensors: filter height $\times$ filter width $\times$
number channels $\times$ number filters. This requires a different
sorting of the components than for fully connected layers. The
procedure is illustrated for two consecutive convolutional layers with
a one-channel $4\times 4$ input and a four-channel $4\times 4$ output
in Figure \ref{fig:conv-attack}. The smallest weights are randomly
distributed over the first half of the \emph{filters}, resulting in a
very likely deactivation of half the channels. 
For each filter of the second layer,
half the \emph{channels} are equipped with the small weights, so that
the negative filter channels are applied to the positive input
channels. The positive filter weights are applied to the deactivated
neurons, and do not contribute to the sum over all channels
for each filter. Thus, deactivation of all
output channels is probable.

Given this layout, we shift the channels of a filter of the second
layer in order not to block the whole network. 
Compared to the previously discussed shifting attack, we have more degrees of freedom: a shift per filter and the number of filters where to apply shifting. The same
 holds for the soft knockout attack, where we
 specify on how many filters in the even layers the permutation is applied.

\textbf{Complexity of Attacks.} The attack's computational complexity is linear in the number of components of the matrices because one pass over them is sufficient for the split into large and small weights.

\subsection{Statistical Analysis of Adversarial Initialization}
The matrices which are permuted in the above attacks are
initialized randomly. To establish that we can expect to observe
a sufficiently large fraction of negative weights, we proceed with a formal
analysis of the statistics of the attacks. The goal is to give
estimates of the probabilities of deactivating certain neurons by
means of adversarial initialization in the above sense. We investigate
how the layer size, the variance of the weights and the magnitude of
the biases influence our attack and show that the input data is indeed
not important for its success. For clarity, we consider the case
of two fully connected layers as presented as the prototype of our
attack. Thus, our architecture is described by
the formula
\[
  \vec y = \relu\big(\mat B\relu(\mat A\vec x + \vec a) + \vec b\big).
\]
Note that the analysis of this case is not merely relevant for
two-layer networks. For the attack it does not matter whether the two
layers are part of a bigger network or not and whether they are the
first layers or somewhere in between other layers, as long as they
interrupt the data flow by deactivating neurons. Additionally, the
analysis of the two fully connected layers basically carries over to
convolutions, the shifting and the soft knockout attack because the
corresponding parameters can be adapted to all cases.

\textbf{Statistics of Adversarial Weights.}
\begin{figure}
  \centering
  \input{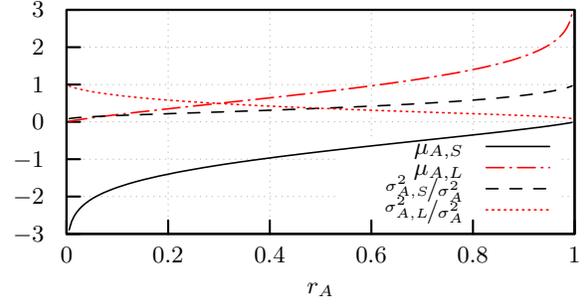}
  \caption{Mean and variance of the weights in the ``small values''
    respectively ``large values'' blocks of $\mat A$.}
  \label{fig:statistics-a}
\end{figure}
As groundwork for the subsequent discussion, we first look at the
statistics of the components of the block matrices $\mat A$ in
equation~\ref{eq:ax}, where the randomly sampled components are split into
two sets of large respectively small values. In particular, we are
interested in the mean values $\mu_{A,S}$ and $\mu_{A,L}$ as well as
the variances $\sigma_{A,S}^2$ and $\sigma_{A,L}^2$ of the components
of the two blocks of $\mat A$, depending on the parameter $r_A$ that
determines the size of the split. The subscript $A$ denotes matrix
$\mat A$, so that we can distinguish the values from those for
$\mat B$ (from equation~\ref{eq:bx}) for which the respective values can be
calculated in a completely analogous way. The quantities that refer to
the block of \emph{small} values have the subscript $S$ and the
respective quantities for the block of \emph{large} values are
sub-scripted with $L$, consistent with the notation in Algorithm
\ref{alg::bleeding}.  We later need the
means and variances for estimating the probability of knocking out
neurons.

We focus on the most relevant case of components that are drawn from a
normal distribution with mean $\mu_A$ and variance $\sigma_A^2$, now
without the subscripts $S$ or $L$ because we refer to the unsplit
values. The distribution of the weights in the ``small values'' block of
$\mat A$ can then be approximated as a normal distribution that is cut
off (i.e. zero for all values greater than some $c$) depending on the
parameter $r_A$ in such a way that the respective part of the original
distribution covers the fraction $r_A$ of the overall probability
mass. Formalizing this, the value of the cut-off-parameter $c$ is
obtained by solving the equation
\begin{equation}
  r_A = \int_{-\infty}^c\frac 1 {\sqrt{2\pi}\sigma_A}\exp\left(-\frac{z^2}{2\sigma_A^2}\right) \diff z
\end{equation}
for $c$. We obtain $c = \sqrt 2 \sigma_A \erf^{-1}(2r_A - 1)$, where
$\erf^{-1}$ is the inverse error function.  As a result, we get the
following probability density distribution for the weights of the
``small values'' block of $\mat A$:
\begin{equation}\label{eq:small-value-pdf}
  f_{A, S}(z) = \begin{cases}\frac 1 {\sqrt{2\pi}\sigma_Ar_A}\exp\left(-\frac{z^2}{2\sigma_A^2}\right) & \text{for}\; z < c, \\
  0 & \text{else.}\end{cases}
\end{equation}
The density $f_{A, L}$ for the ``large values'' block is
found accordingly.

\begin{figure*}
  \centering
  \input{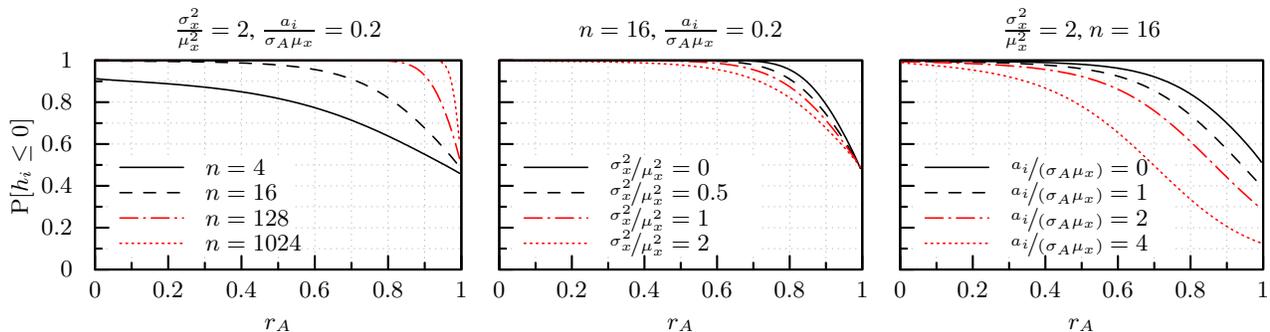}
  \caption{Illustration of the probability of obtaining deactivated
    neurons after the first layer, depending on the relative block
    size $r_A$ and selected values for the other parameters.}
  \label{fig:geq-zero-prob}
\end{figure*}

Before proceeding, we introduce the shorthand notation
\begin{equation}
  g(r) := \sqrt{\pi}\exp\left(\big(\erf^{-1}(2r-1)\big)^2\right),
\end{equation}
which will prove useful for presenting the results in a more succinct
form. From equation~\ref{eq:small-value-pdf} a straightforward integration
yields
\begin{equation}\label{eq:mua}
    \mu_{A,S} = -\frac {\sigma_A} {\sqrt{2}r_Ag(r_A)}, \,
    \mu_{A,L} = \frac {\sigma_A} {\sqrt{2}(1-r_A)g(r_A)}.
\end{equation}
Likewise, the variances of the components of the blocks are:
\begin{subequations}\label{eq:sigmaa}
  \begin{align}
    \sigma_{A,S}^2 &= \sigma_A^2 + \sqrt{2}\sigma_A\erf^{-1}(2r_A - 1)\mu_{A,S} -  \mu_{A,S}^2 \\
    \sigma_{A,L}^2 &= \sigma_A^2 + \sqrt{2}\sigma_A\erf^{-1}(2r_A - 1)\mu_{A,L} -  \mu_{A,L}^2
  \end{align}
\end{subequations}
The means and variances are plotted in Figure~\ref{fig:statistics-a}. In our
model, $\mu_{A,S}$ is always negative while $\mu_{A,L}$ is always
positive because there is always an imbalance between positive and
negative values. Large or small values of $r_A$ make the statistics
of the larger block look like those of the original matrix $\mat A$,
while the few values in the small block have a mean with large
absolute value and small variance.

\textbf{First Layer.}
With these results in mind, we are ready to analyze the effect of the
first layer of equation~\ref{eq:consecutive-layers} with a weight matrix
$\mat A$ that is split according to equation~\ref{eq:ax} and a bias $\vec a$.
With the convenient definition
\begin{equation}
  \vec h = \mat A \vec x + \vec a
\end{equation}
we can estimate the expected value $\mu_{h,i} := \E[h_i]$ of the components of
$\vec h$ given random inputs and fixed weights and biases. We define 
the expected values $\mu_{x} := \E[x_i] $ (for
any $i$, see below) as well as $\mu_{A,i} := \E[A_{i:}]$ and get
\begin{equation}\label{eq:muh}
\mu_{h,i} = \!\! \sum_{j=1}^n \! \! A_{ij}\E[x_j] + a_i
\approx n\mu_x \frac 1 n \sum_{j=1}^n A_{ij} + a_i  \approx n\mu_x \mu_{A,i}
\end{equation}
The first approximation is based on the premise that the components of
$\vec x$ are approximately equally distributed while the second
approximation gets better with increasing $n$. The assumption of equal
distributions is particularly justified if the first layer of our
model is not the first layer of the network because in that case input
differences are evened out by forming sums with random weights in the
previous layers. If $\vec x$ is actually the input layer, we can of
course not always guarantee a particular distribution of its
components. Nevertheless, given typical actual datasets, it is still
reasonable to assume similar distributions for a sufficiently large part of
the features so that the approximation is meaningful.

Under the same assumptions and with the variance $\sigma_{A, i}^2$ of
the elements of the $i$-th row of $\mat A$ as well as the variance
$\sigma_x^2$ of the components of $\vec x$, together with the premise
that the components of $\mat A$ and those of $\vec x$ are
statistically independent, we obtain:
\begin{equation}
  \begin{split}
    \E[h_i^2] &\approx \E[x]^2n(n-1)\mu_{A,i}^2 + 2a_i n \E[x]\E[A_{i:}] \\
    &+ \E[x^2]n\big(\sigma_{A, i}^2 + \mu_{A,i}^2\big) + a_i^2
  \end{split}
\end{equation}
With that, we get the variance of $h_i$:
\begin{equation}\label{eq:sigmah}
  \sigma_{h,i}^2 := \E[h_i^2] - \E[h_i]^2 \approx n\left(\mu_{A,i}^2\sigma_x^2 + \sigma_{A,i}^2\sigma_x^2 + \sigma_{A,i}^2\mu_x^2\right)
\end{equation}
As we assume $n$ to be large enough for our approximations to be
reasonable, we can apply the central limit theorem that tells us that
$h_i$ will approximately follow a normal distribution
$\mathcal{N}(\mu_{h,i}, \sigma_{h,i}^2)$. Because of this,
equation~\ref{eq:muh} and equation~\ref{eq:sigmah} completely determine the
distribution of $h_i$ and the probability for $h_i$ to be smaller than
or equal to zero is readily estimated as
\begin{equation}\label{eq:zero-prob}
  \prob[h_i \leq 0] = \int_{-\infty}^0 \!\!\!\!\!\!\!\!\mathcal{N}(h; \mu_{h,i}, \sigma_{h,i}^2) \diff h = \frac 1 2 - \frac 1 2 \erf\left(\frac {\mu_{h,i}} {\sigma_{h,i}\sqrt{2}}\right).
\end{equation}

For normally distributed weights, equation~\ref{eq:muh} and equation~\ref{eq:sigmah}
can be calculated on the basis of our previous results for the
statistics of $\mat A$, given in equation~\ref{eq:mua} and
equation~\ref{eq:sigmaa}. Under our assumptions, the row index $i$ matters
only in so far that it either belongs to the (hopefully) deactivated
neurons or to the other block. We find that $\nicefrac{ \mu_{h,S}} { \sigma_{h,S}}$ equals

{\footnotesize
  \begin{subequations}
    \begin{equation}
      \frac { \frac {2} {\sqrt{n}}\left(\frac {a_i} {\sigma_A\mu_x}\right) r_Ag(r_A)- \sqrt{\frac n 2}}{\sqrt{\left(r_A^2 g(r_A)^2 - r_A\erf^{-1}(2r_A-1)g(r_A)\right)\left(\frac {\sigma_x^2}{\mu_x^2} + 1\right)- \frac 1 2}}.
    \end{equation}
{\normalsize The analogous expression for $\nicefrac {\mu_{h,L}} {\sigma_{h,L}}$ with $\scriptstyle \bar{r}_A=1-r_A$ is}
    \begin{equation}
      \frac {\frac{ 2} {\sqrt{n}}\left(\frac {a_i} {\sigma_A\mu_x}\right)\bar{r}_A g(r_A) + \sqrt{\frac n 2}}{\sqrt{\left(\bar{r}_A^2 g(r_A)^2 + \bar{r}_A\erf^{-1}(2r_A-1)g(r_A)\right)\!\left(\frac {\sigma_x^2}{\mu_x^2} + 1\!\right)\!-\!\frac 1 2}}.
    \end{equation}
  \end{subequations}}
Together with equation~\ref{eq:zero-prob} we obtain estimations for the
probabilities of switching off neurons after the first layer. The
behavior depends on three dimensionless\footnote{This concept of
  ``dimensionless'' stems from physics and related disciplines, where
  similar quantities are used to describe and classify complex systems
  in a unit-independent way.} parameters that are given due to the
setup: The input dimension $n$, the ratio
$\nicefrac {a_i} {\sigma_A\mu_x}$ that corresponds to the relative
importance of the bias and $\nicefrac {\sigma_x^2}{\mu_x^2}$, which can
roughly be described as a measure of sharpness of the input
distribution. The influence of these parameters can be observed in
Figure~\ref{fig:geq-zero-prob}. As expected, a significant positive
bias deteriorates the probability; nevertheless it must be unusually
high to have a significant effect. For large $n$, the probabilities
are more distinct because the statistics get sharper. The
characteristics of the input data, on the other hand, do not play a
big role, as it can be seen in the second diagram. Note that the
variance of the weights does not directly influence the
probabilities. Overall we can conclude that the chances of
deactivating neurons is indeed high for realistic choices of
parameters and that the characteristics of the input data hardly
influence the system.

\textbf{Second Layer.} The statistical analysis of the effect of the
second layer is very similar to that of the first layer, just
significantly more complex in terms of the length of the expressions
and cases that have to be distinguished. As there is not much to learn
from that, we leave out the details of the respective computation and
simply remark that after the second layer neurons are indeed
deactivated with a high probability for realistic parameters.

 
\section{Empirical Evaluation}\label{sec::setup}
We now evaluate the previously derived attacks. Before we present our
results, we detail the setting, describe the datasets and
architectures we use and explain how we illustrate and plot our findings.

\newcommand\clhint[4]{\tikz{\definecolor{mycl}{RGB}{#1, #2, #3}; \node[fill=mycl, inner sep=1pt, rounded corners=0.5pt, minimum width=4em] at (0, 0) {#4};}}

\begin{table}[t]
  \footnotesize
  \centering
	\caption{Overview of datasets used.}\label{table:datasets}
	\begin{tabular}{@{}l rrrrrr@{}}
	    \toprule[1.5pt]
		Name & number of & number of  & random  & assigned\\
		& features & samples  & guess  & color\\ 
        \midrule
        MNIST & $28\! \times\! 28\! \times\! 1$  & $70\,000$& $10$\%  & not plotted \\
        F-MNIST & $28\! \times\! 28\! \times\! 1$ & $70\,000$& $10$\% & \clhint{255}{223}{143}{yellow}  \\
        CIFAR10 & $32\! \times\! 32\! \times\! 3$ & $60\,000$ & $10$\% & \clhint{255}{161}{127}{orange} \\
		\bottomrule[1.5pt]
	\end{tabular}
\end{table}

\textbf{Datasets.} We evaluate the attacks on a range of datasets,
which are summarized in Figure~\ref{table:datasets}. We  
classify middle-sized tasks;
MNIST~\cite{lecun1998gradient} and the more challenging
Fashion-MNIST~\cite{xiao2017/online}. Both consist of black and white
pictures of size $28 \times 28$ pixels. The former dataset contains
the handwritten digits 0-9, the latter images of clothing such as
shoes, hats, or trousers. Finally and as a more challenging task, we
choose the classification of images from the
CIFAR10~\cite{krizhevsky2009learning} dataset. This dataset consists
of small, colored images (sized $32 \times 32$ pixels) of trucks,
cars, planes etc.

\textbf{Architectures.} We evaluate two different kinds of
architectures, fully connected networks and convolutional
networks. All our fully connected networks contain $\nicefrac n 2$
neurons in the first hidden layer, where $n$ is the number of
features. The second hidden layer 
has 49 neurons for the two MNIST tasks. As an
example for a convolutional architecture, we use LeNet on CIFAR10~\cite{lecun1998gradient}.

All networks are initialized using the He initializer and constant bias.
The fully connected networks are trained for
300 epochs on both MNIST variants. 
LeNet is trained for 200 epochs. We initialize all
networks using the He initializer and optimize them with the Adam
optimizer with its default learning rate of 0.001. 
However, Appendix~\ref{app:altattacks}-\ref{app:lr}, we show  that 
initializer, optimizer, learning rate and even activation function 
do not prevent vulnerability.

\begin{figure}[t]
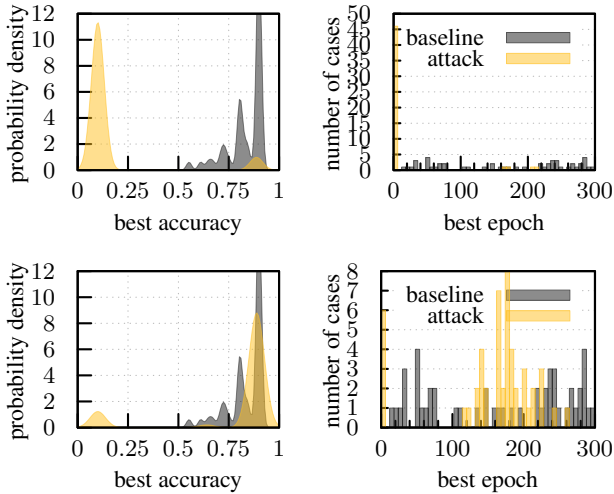

  \input{gp/FMNIST_he_Poison_57}
  \input{gp/FMNIST_he_Poison_60}
  \caption{The soft knockout attack allows little control over the
    networks accuracy: Fashion-MNIST, fully connected network,
    $r=0.25$ (upper) versus $r=0.2$ (lower). The networks either fail
    entirely or converge slower.}
    \label{fig::bleeding_details}
\end{figure}

   \begin{figure}[t]
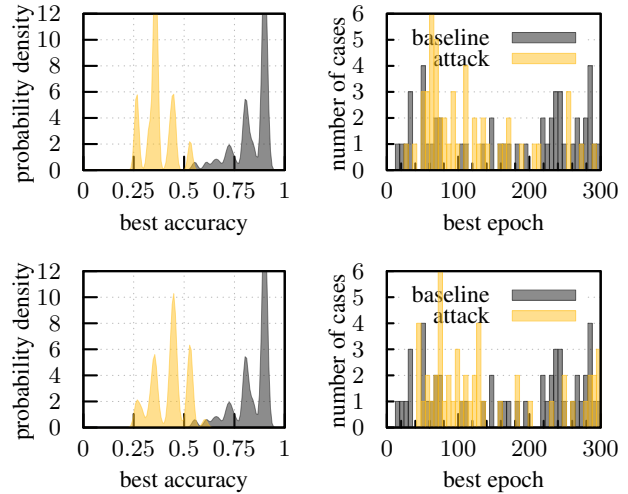

  \input{gp/FMNIST_he_Poison_58}
  \input{gp/FMNIST_he_Poison_59}
  \caption{The shift attack on the Fashion-MNIST task. Upper plot set shift to four, lower to eight.}
    \label{fig::shiftDense}
\end{figure}

\textbf{Presentation of results.} We are interested in how our attacks
affect the probability to get a well performing network after
training. Towards this end, we mainly consider two quantities: the
best accuracy that is reached during training and the epoch in which
it has been reached. Due to the random initialization and the way in
which neural networks work and are trained, there is not a single best accuracy and a
particular best epoch for a given task, but a distribution over
accuracies and epochs over different seeds for the same architecture. We approximate these distributions by
evaluating a sample of 50 networks with different seeds for the random
initializer\footnote{We keep the same 50 seeds in all
  experiments for comparability. However, due to effects from
  parallelization on GPUs, the accuracy might differ by up to
  2\% for seemingly identical setups.}. We plot the smoothed
probability density function over the best test accuracies during training
and the epochs at which this accuracy was observed. While we use
Gaussian kernel density estimation for the former, the latter is
depicted using histograms. Both distributions are compared to a
baseline derived from a sample of 50 networks with the same 50
random seeds, trained without adversary.

\subsection{Soft Knockout Attack}

\begin{figure*}[t]
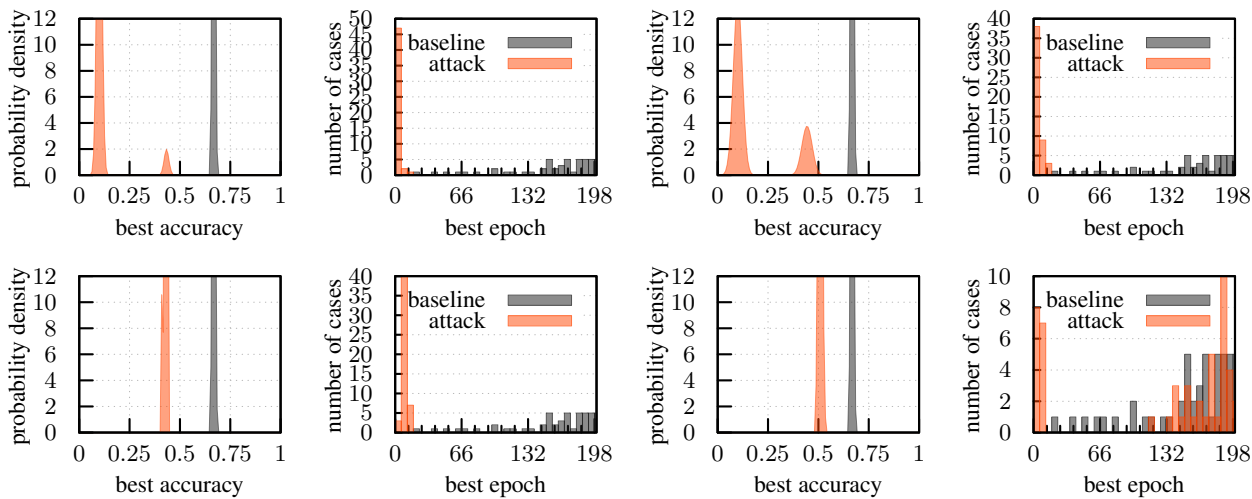

\begin{subfigure}[b]{0.49\linewidth}
  \input{gp/shift-experiments2_cifar10_1_4}
  \input{gp/shift-experiments2_cifar10_1_8}
\end{subfigure}	
\begin{subfigure}[b]{0.49\linewidth}
  \input{gp/shift-experiments2_cifar10_16_4}
  \input{gp/shift-experiments2_cifar10_16_8}
\end{subfigure}
  \caption{The shift attack on convolutional architectures. We vary both the shift (upper plots' shift is four, lower plots' shift is eight) and the number of filters the shift is applied to (left plots' is applied to one filter, right ones to sixteen). Evaluation dataset is CIFAR10.}
    \label{fig::shiftConv}
\end{figure*}

For the soft knockout attack, we
control the size of the split between small and large values of the
weight matrices in order not to knock out all the neurons at once.
The experiments show that this gives little control over the
performance of a network: On fully connected networks, training either
fails entirely, or the network achieves normal accuracy (however after
a larger number of epochs). We conclude that as soon as the networks have some non-vanishing chance
of updating the weights (which is the idea of a soft knockout), they
can recover from the bad initialization.

We plot the
results on Fashion-MNIST in Figure~\ref{fig::bleeding_details}. We depict
the results for $r=0.2$ and $r=0.25$. A parameter $r>0.3$ leads to
complete failure to learn: all accuracies are equivalent to
guessing. We observe that networks that perform as good as random guess
usually perform best in their first iteration, and do not improve
during training or more concretely, they do not train at all. This 
is visible as well for $r=0.25$ and hence in the
upper plot of Figure~\ref{fig::bleeding_details}. We picked Fashion-MNIST to
illustrate this, although it occurs in general. For slightly lower
$r=0.2$, however, most seeds achieve baseline accuracy. Once again the
training time is increased on average, as visible in
Figure~\ref{fig::bleeding_details}.

We finally apply an adapted version of the soft knockout attack to a
convolutional network on CIFAR10. 
We set in every second (uneven) layer
 to $r=0.2$, otherwise to $0.5$. We compare between applying 
 softening to only one
filter or to sixteen filters (the latter means half of the available
filters). In contrast to the results for fully connected networks, we
do observe a reasonable decrease in the best achieved accuracy, which 
is now less than 50\%. This
accuracy is typically reached at the beginning of the training,
meaning that the networks actually get worse during training instead of
converging to a good configuration. We also ran the experiments on very 
small datasets confirming the our results
 in the appendix.

\subsection{Shift Attack}

The shift attack gives more
fine-grained control over the network that the victim trains. For
fully connected networks, the shift parameter is equivalent to the
number of active neurons in the network. Our experiments show that 
a umber of 10 (MNIST)/ 12 (Fashion MNIST) neurons suffices to learn 
the task with unchanged accuracy.  We thus depict our results with a
shift parameter of 4 and 8 on Fashion-MNIST in
Figure~\ref{fig::shiftDense}. The accuracy is decreased significantly but
the network does not fail completely.

As expected, as the shift decreases and less neurons are available to
the network, the networks' performance decreases. On
Fashion-MNIST, we observe an increase in training time of around $50$
epochs. This is less clear for MNIST, where several networks are
failing, and achieve their best (random guess) accuracy in epoch
one. The plots for MNIST can be found in the appendix. 

We additionally depict the results on convolutional networks on the
CIFAR10 dataset in Figure~\ref{fig::shiftConv}. We once again apply a shift
of either four or eight and set the number of filters this shift is applied
to to either one or sixteen. As for the fully connected networks, we
observe a strong decrease in accuracy.
We observe that most networks fail for a shift of four, independently
of the number of filters that are affected. With a shift of eight, the
networks obtain on average an accuracy around 43\% if one filter is
affected and around 50\% if the number of filters is increased to 16.
In contrast to the previous attacks on dense networks, we mostly
observe a decrease in training time. An exception to this is a shift
of four applied to sixteen filters, where the training time is either very
short or rather long. Further experiments concerning the inference of learning rate, optimizer and initializer 
are in the appendix.

\section{Why would I care?}
First, one might wonder how an attacker might even be able to alter the code of the library. The idea that trusting libraries can be recognized as a threat \cite{grace2012unsafe,backes2016reliable,lauinger2018thou}
has also been recognized in ML~\cite{liu2017fault,DBLP:conf/sp/XiaoLZX18}.
A simple drive-by download is enough to infect a machine with the malicious code~\cite{DBLP:conf/auisc/LeWGK13},
if no corresponding defense is in place~\cite{canali2011prophiler,DBLP:conf/uss/KapravelosSCKV13,javed2019prediction}.

\begin{figure}[t]
  \input{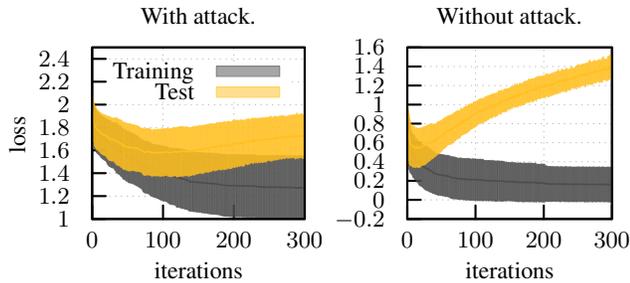}
  \caption{Loss during training on Fashion MNIST (fully connected network, shift is 4). Along with the achievable accuracy, the scale of the loss is unknown to the victim.}\label{fig::losses}
\end{figure}

Furthermore, one might ask whether a user would actually fall for such an easy-to-fix attack as maliciously permuted weights. We argue that this hinges on the user's awareness of the attack and that current debugging routines hardly take initialization into account. In order to underpin this statement, we carry out a study on \url{stackoverflow.com} and \url{stackexchange.com}, popular and typical Q\&A sites for programming-related issues.
We browse the replies to questions concerning neural network failure and check whether people would discover our attack based on this advice (the full study can be found in Appendix~\ref{app:study}).
In a nutshell, for the specific setting the attack causes,
in 115 relevant questions, the majority of the answers either point out a bug (32.2\%), concern the data (31.3\%), or suggest altering the model (30.4\%).
In only 3.5\% (i.e.\ four) of the cases the suggestions could give away the attack.
However, in three of these cases, the setting described by the user suggests that the model is not learning at all, or the loss is severely diverging. For our attack, the loss does not look that suspicious, as can be seen in Figure~\ref{fig::losses}. This leaves \emph{one} answer that would actually point into the direction of our attack for the symptoms it causes: ``\emph{Gradient check your implementation with a small batch of data and be aware of the pitfalls}''
This is still far from a direct hint. Overall, we conclude that there is a lack of awareness on the importance of the initial weights.

\section{Conclusion}
In this paper, we show that the threat of adversarial initialization
goes far beyond previously known attacks that induced overfitting.
The studied attacks permute the initial weight matrices
before training.  On the MNIST benchmark, benign accuracy is
easily $>98$\%. Under the attack, the victim is unable to achieve an
accuracy $>50$\%. On Fashion MNIST, the attacker limits the accuracy
from $>90$\% to around $50$\%.
We further show that the loss is unsuspicious, and that the user, given current
knowledge, will not discover the source of the bad performance.
 In addition to these
empirical results, we formally derive statistical evidence that the
attacks succeed for standard initializations and is 
independent of
the input distribution and thus the task at hand.

\section*{Acknowledgments}
This work was supported by the German Federal Ministry of Education and
Research (BMBF) through funding for the project CISPA\_AutSec (FKZ: 16KIS0753). Marius Mosbach 
acknowledges partial support by the German Research Foundation (DFG) 
as part of SFB 1102.

\bibliographystyle{plainnat}
\small
\bibliography{main}

\normalsize
\clearpage
\appendix

\section{Additional datasets used in the Appendix}
For completeness, we introduce all datasets, also the datasets introduced
in the main paper. We evaluate on a range of datasets,
which are summarized in Table~\ref{table:datasetsApp}. We choose two small
datasets, spam~\cite{Lichman:2013} and credit~\cite{Lichman:2013}. The
spam dataset defines a binary classification task. Based on 56 binary
or real valued features, emails are to be classified as ``ham'' or
``spam''. Credit contains $14$ features and $690$ instances, and our
task is to predict whether an applicant is granted a credit demand.
We furthermore consider classification tasks on middle-sized datasets,
MNIST~\cite{lecun1998gradient} and the more challenging
Fashion-MNIST~\cite{xiao2017/online}. Both consist of black and white
pictures of size $28 \times 28$ pixels. The former dataset contains
the handwritten digits 0-9, the latter images of clothing such as
shoes, hats, or trousers. Finally and as a more challenging task, we
choose the classification of images from the
CIFAR10~\cite{krizhevsky2009learning} dataset. This dataset consists
of small, colored images (sized $32 \times 32$ pixels) of trucks,
cars, planes etc.

\begin{table}[t]
  \small
  \centering
	\caption{Overview of datasets used.}\label{table:datasetsApp}
	\begin{tabular}{@{}l rrrrrr@{}}
	    \toprule[1.5pt]
		Name & \# of & \# of  & random & kind of & assigned\\
		& features & samples  & guess & features & color\\
        \midrule
        Credit & $14$ & $690$& $60$\% & mix & \clhint{223}{191}{255}{purple} \\
        Spam & $56$ & $4\,601$ & $70$\% & mix & \clhint{194}{230}{245}{blue} \\ 
        MNIST & $784$  & $70\,000$& $10$\% & real & \clhint{150}{196}{170}{green} \\
        F-MNIST & $784$ & $70\,000$& $10$\% & real & \clhint{255}{223}{143}{yellow}  \\
        CIFAR10 & $3072$ & $60\,000$ & $10$\% & real & \clhint{255}{161}{127}{orange} \\
		\bottomrule[1.5pt]
	\end{tabular}
\end{table}
\section{Shift Attack on MNIST}
\begin{figure}[t]
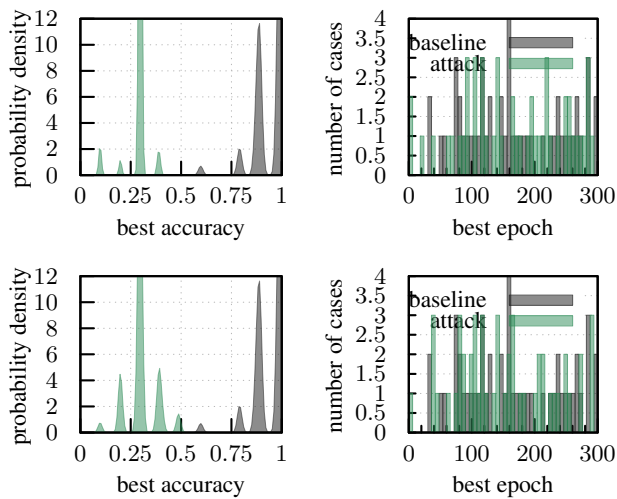

  \input{gp/MNIST_he_Poison_58}
  \input{gp/MNIST_he_Poison_59}
  \caption{The shift attack on the Fashion-MNIST task. Upper plot set shift to four, lower to eight.}
    \label{fig::appshiftDense}
\end{figure}

We depict the results on MNIST which are described in the main paper in Figure~\ref{fig::appshiftDense}.

\section{Evaluation on Alternative Attacks.}\label{app:altattacks}
For the sake of completeness, we present 
some alternative attacks in this section. This evaluation
follows the set up described in the main paper. The results
presented here can be compared against those for our main attacks.

\begin{figure}[!hb]
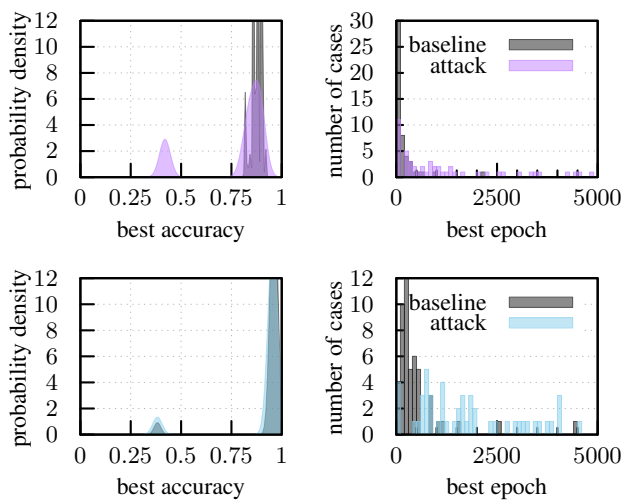

  \input{gp/Credit_he_Poison_35}
  \input{gp/SPAM_he_Poison_35}
  \caption{Setting variance to non-ideal value on the credit (above)
    and the spam (below) task.}
  \label{figAp::alterVar}
\end{figure}

For the first attack we change the variance of the weights: instead of
offsetting it to the ideal value
$\nicefrac 2 {\mathrm{fan}_\mathrm{in}}$, we set it to
$\nicefrac 2 {\mathrm{fan}_\mathrm{out}}$. We report the results of our
experiment on the Credit and Spam tasks in figure~\ref{figAp::alterVar}. We
observe an increase in training time, and an increase of
non-converging networks on the credit data. The accuracy on the spam
data does not change.

\begin{figure}[!hb]
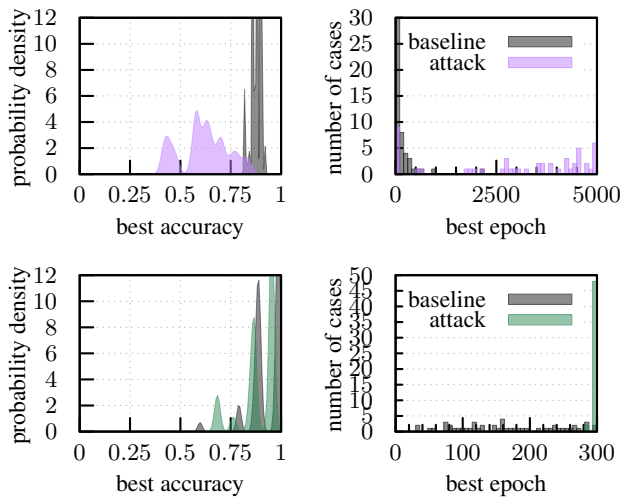

  \input{gp/Credit_he_Full_1e-06}
  \input{gp/MNIST_he_Full_1e-06}
  \caption{Maliciously set learning rate on credit (above) and MNIST
    (below) task.}\label{figAp::learningRate}
\end{figure}

In a second experiment, we alter the leaning rate maliciously to slow
down learning. Instead of the default $10^{-3}$, we set the learning
rate to $10^{-6}$ and depict the results in
figure~\ref{figAp::learningRate}. We observe both intended effects, as the
training time increases and the accuracies decrease.

\begin{figure}[!hb]
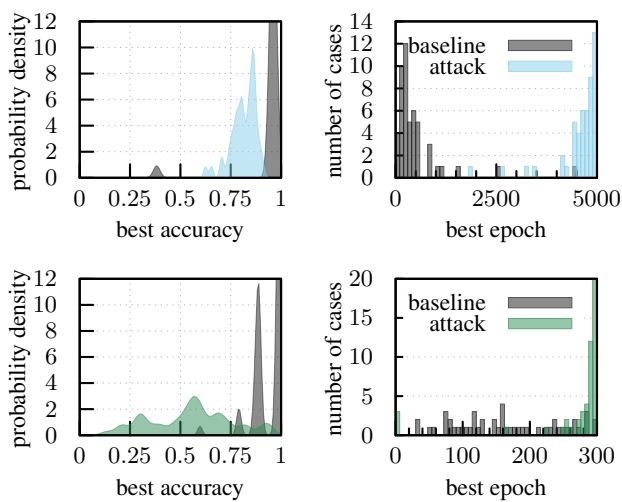

  \input{gp/SPAM_he_A0.005}
  \input{gp/MNIST_he_MNISTdropA0.01}
  \caption{Maliciously set dropout (attack) and benign training
    (baseline) on spam (above) and MNIST
    (below).}
  \label{figApp::dropoout}
\end{figure}

Finally, we consider the effect of choosing a very large dropout
probability during training. We evaluate this on Spam (dropout rate
0.005) and MNIST (dropout rate 0.01). We depict the results of our
experiments in Figure~\ref{figApp::dropoout}. We again observe both desired
properties: an increase in training time and a decrease in accuracy.

\section{Optimization based attack for small networks}\label{sec:small-networks}
Taking into account our formal analysis and the dependence on the
layer size, we see that small networks (in our case the networks
trained on credit and spam data, see below) are not very susceptible to the
attacks that were discussed so far. Their weight matrices are so small
that the statistics are not sharp enough for guaranteeing deactivated
neurons with a sufficient probability, 
rendering the overall scheme useless.

However, those networks can be targeted as well. We formulate the
following general knockout optimization problem, where our network $F$
is parametrized with the weight matrices $\mat{W}_i$:
\begin{equation}
  \min_{\{\mat{W}_i\}} \left(\sum_c\sum_jF_c(\vec{X}_j)\right),\quad\text{s.t.}\; \|\mat{W}_i\|_\mathrm{F} = \text{const.}
\end{equation}

This expression describes a minimization of the output of the last
layer for each class $c$. The constraint keeps the Frobenius norm of the weights constant
so that they usually tend to stay close to the original weights,
making the attack stealthy.  While this problem is formulated in a way
that requires full knowledge of the network and the data, we can
obtain reasonable results on a batch of data drawn from a uniform
distribution and replacing later parts of the network with randomly
drawn fresh matrices.

We mention this attack for the sake of completeness. In practice,
small networks are not as relevant as the large ones, so that a
failure on them is not problematic. This alternative
approach underlines once more
 that our attacks are merely specific instances of a larger
class of initialization attacks. 
We implemented this attack using an
off-the-shelf optimizer provided by Scipy~\cite{scipy}. 

\begin{figure}[!h]
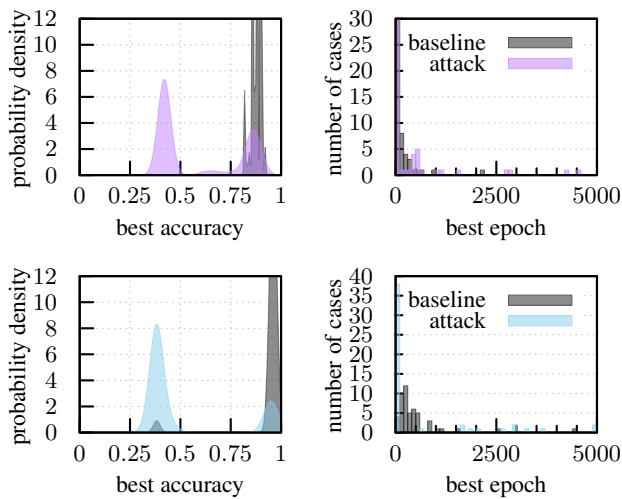

  \input{gp/Credit_he_Poison_43}
  \input{gp/SPAM_he_Poison_43}
  \caption{Using off-the-shelf optimization to knock out neurons on
    the credit (upper) and spam (lower) tasks.}
    \label{fig::att43}
\end{figure}

The optimization finishes quickly (runtime $<1$s) for small
networks (on credit and  spam data). Each of this networks has in the second and third layer half as many neurons as features, e.g. 7 for credit and 28 for spam. We depict the resulting accuracy
distributions in Figure~\ref{fig::att43}. Independent of the dataset,
networks either fail completely (best accuracy in iteration $0$), or
 converge to the original accuracy (however slower). We conclude
that small networks can be targeted, albeit it is hard to decrease
accuracy in a stealthy way.

\section{Attacking Sigmoid Activations}\label{app:sigmoid}
We have defined particular attacks on the ReLU activation. Here, we want to briefly describe an attack on sigmoid networks as a proof of concept. In contrast to ReLU, we cannot not knock out neurons, and thus aim to trigger vanishing gradients.

We formalize the corresponding algorithm of the attack in algorithm~\ref{alg::extreme}. The idea is to set all elements very small, with a single row of large values that fixes mean and variance as in the benign weight matrix. We start with a template matrix that is a zero matrix and contains randomly drawn zeros and ones in the first column. This matrix is adjusted to the size of the target weight matrix (line 2
). We then compute the difference in mean and variance (line 3 
and 4
). These differences help us to re-scale the values as to mimic the original statistics (line 5 
 and following). We first alter all zero weights to be able to obtain reasonable variance, then adapt the variance and correct again for the mean.

\begin{algorithm}
  \caption{\textbf{Vanishing Weights Attack.} Given a stream of weights
    $\mathcal{W} = \{\mat{W}_1, ...\}$ and a template matrix, output
    altered weights that will impede training.}\label{alg::extreme}
  \begin{algorithmic}[1]
    \REQUIRE $\mathcal W$, template
    \FOR{$\mat{W}_i \in \mathcal W$}
    \STATE $\mat{W}_{new} \leftarrow $ reshape$(\mat{W}_i$, template$)$\label{algo:reshape}
    \STATE $\mat{mean} \leftarrow$ mean$(\mat{W}_i)-$mean$(\mat{W}_{new})$ \label{algo:eMean}
    \STATE $\mat{var} \leftarrow$ variance$(\mat{W}_i)/$variance$(\mat{W}_{new})$ \label{algo:eVar}
    \STATE $\mat{W}_{new} \leftarrow \mat{W}_{new} + \mat{mean}$ \label{algo:adjust}
    \STATE $\mat{W}_{new} \leftarrow \mat{W}_{new} * \sqrt[2]{\mat{var}}$
    \STATE $\mat{W}_{new} \leftarrow \mat{W}_{new} + \mat{mean}$
    \STATE $\mat{W}_{i} \leftarrow \mat{W}_{new}$
    \ENDFOR
  \end{algorithmic}
\end{algorithm}

\begin{figure}
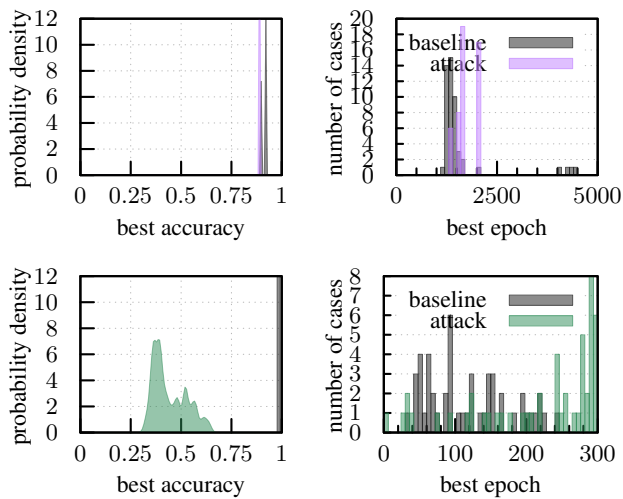

  \input{gp/Credit_normal_Poison_101}
  \input{gp/MNIST_he_Sig_Poison_101}
  \caption{The attack on sigmoid networks on the Credit (above) and MNIST data (below).}
    \label{fig::sigattack}
\end{figure}

We now empirically evaluate this attack, once again following the set up described
in the main paper. The results are depicted in Figure~\ref{fig::sigattack}. We observe that accuracy is consistently decreased, and the training time generally increases. More specifically, on Credit, the accuracy is generally $0.9$, and the benign accuracy slightly higher. The training time increases slightly from roughly 1200 epochs till up to 2000 epochs until convergence. On MNIST, the increase in training time is less clear. Yet, the decrease in accuracy is very strong, and the maximum accuracy observed is slight below $0.7$.

\begin{figure}
  \input{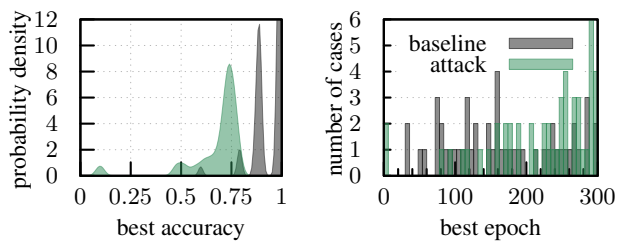}
  \caption{The sigmoid attack evaluated using ReLU activations on MNIST data.}
    \label{fig::sigattackREL}
\end{figure}

In general, vanishing gradients can also occur in ReLU activations. We thus evaluate the attack on MNIST using ReLU activations and depict the results in Figure~\ref{fig::sigattackREL}.
We observe a decrease in accuracy (limited by roughly $0.8$\%) and a slight increase in epochs trained.

\section{Evaluation of Initializer and Optimizer.}\label{app:sgdGlorot}
In the main evaluation, we focus on He initialization and the Adam
optimizer. In this appendix, we add additional results using the
Glorot initializer and the SGD optimizer to show that the difference
in vulnerability is negligible.  We again follow the set up described
in the main paper.

\begin{figure*}[!h]
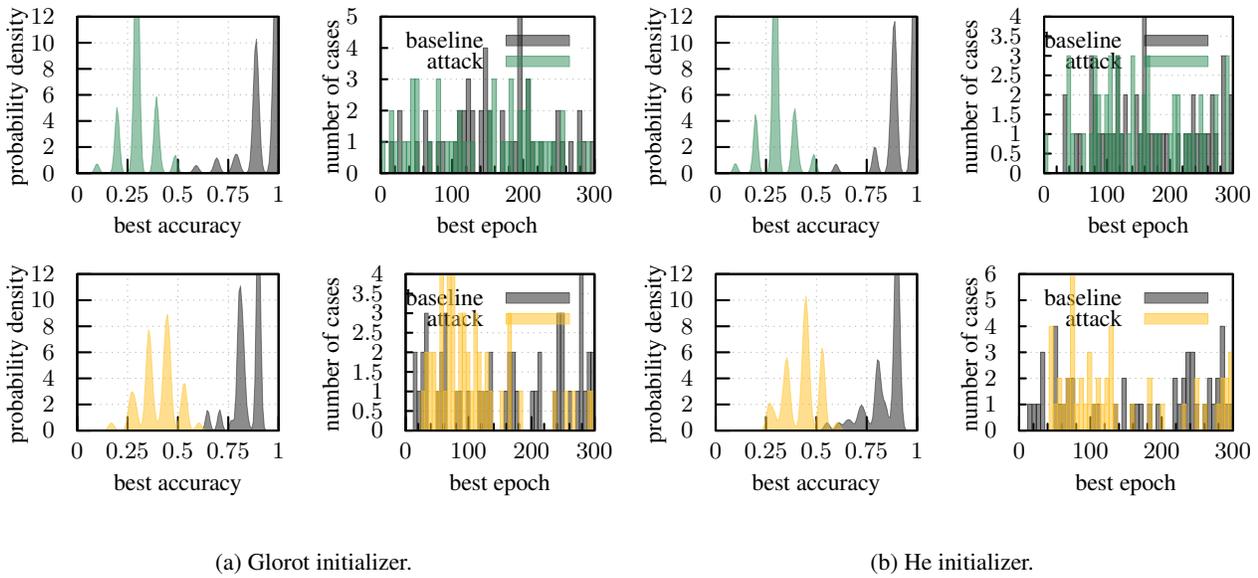

\begin{subfigure}[h]{0.49\linewidth}
  \input{gp/MNIST_glorot_Poison_59}
  \input{gp/FMNIST_glorot_Poison_59}
  \caption{Glorot initializer.}
\end{subfigure}
\begin{subfigure}[h]{0.49\linewidth}
  \input{gp/MNIST_he_Poison_59}
  \input{gp/FMNIST_he_Poison_59}
    \caption{He initializer.}
\end{subfigure}
\caption{The influence of the initializer on vulnerability to the
  shift attack. Datasets are MNIST (above) and Fashion-MNIST (below),
  shift is set to $8$.}
    \label{fig::shifHevsGlo}
\end{figure*}

First, we evaluate the initializer: we study the effect of the shift attack on
MNIST and Fashion-MNIST using the Glorot initializer in
Figure~\ref{fig::shifHevsGlo}. He performs better than the Glorot
initializer under and without the attack. Given that the attacks uses
negative weights, and that both He and Glorot initialize from a (slightly differently)
truncated normal, these results can be excepted.

\begin{figure*}[!h]
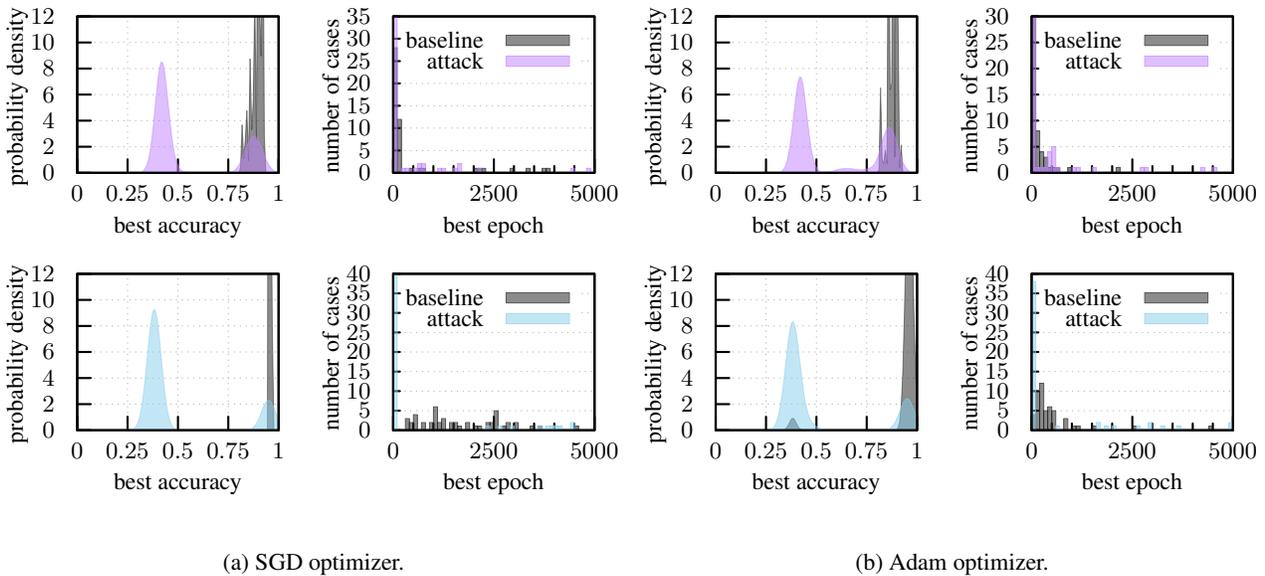

\begin{subfigure}[b]{0.49\linewidth}
   \input{gp/SGD_Credit_he_Poison_43}
  \input{gp/SGD_SPAM_he_Poison_43}
  \caption{SGD optimizer.}
\end{subfigure}
\begin{subfigure}[b]{0.49\linewidth}
   \input{gp/Credit_he_Poison_43}
  \input{gp/SPAM_he_Poison_43}
    \caption{Adam optimizer.}
\end{subfigure}
\caption{The influence of the optimizer on vulnerability to the
  optimization attack. Datasets are credit (above) and spam (below).}
    \label{fig::optSGDvsAdam}
\end{figure*}

Additionally, we compare SGD and Adam optimizer on the optimization attack in
Figure~\ref{fig::optSGDvsAdam}. Adam converges earlier on both datasets and
yields higher accuracy both without and under attack. This results are
to be expected as well: the optimizer can only slightly affect the
gradient that is computed. When inputting only negative weights, however,
the resulting gradient will always be zero. In particular for the shift attack,
the observed difference
between Adam and SGD thus stems purely from their ability to train
the remaining part of the network.

We conclude that the attacks do not rely on a particular
initializer or optimizer.

\section{Defense via Learning Rate Calibration}\label{app:lr}
A victim of our attacks might be tempted to reconfigure the learning rate
to obtain better results with the capacity reduced networks. We thus experiment on
the influence of the learning rate.
We depict four learning rates, $0.01$, $0.005$, and $0.0005$ as compared
to the default $0.001$ studied in the main experiments. We depict the results in \ref{alg::extreme}
ordered from the largest learning rate above, the smallest below. In general,
the learning rate does not alleviate the effect of the attack. Larger learning
rates allow to improve on the original, low accuracy, however still only yield results
around $0.5$. In general, more networks fail completely with
higher learning rates. A smaller learning rate leads to faster convergence,
where however the decreased
accuracy does not change. The best
observed accuracy increases from $40$\% to $55$\%, and is thus still very much below
the best benign accuracy of $0.99$.

\begin{figure}
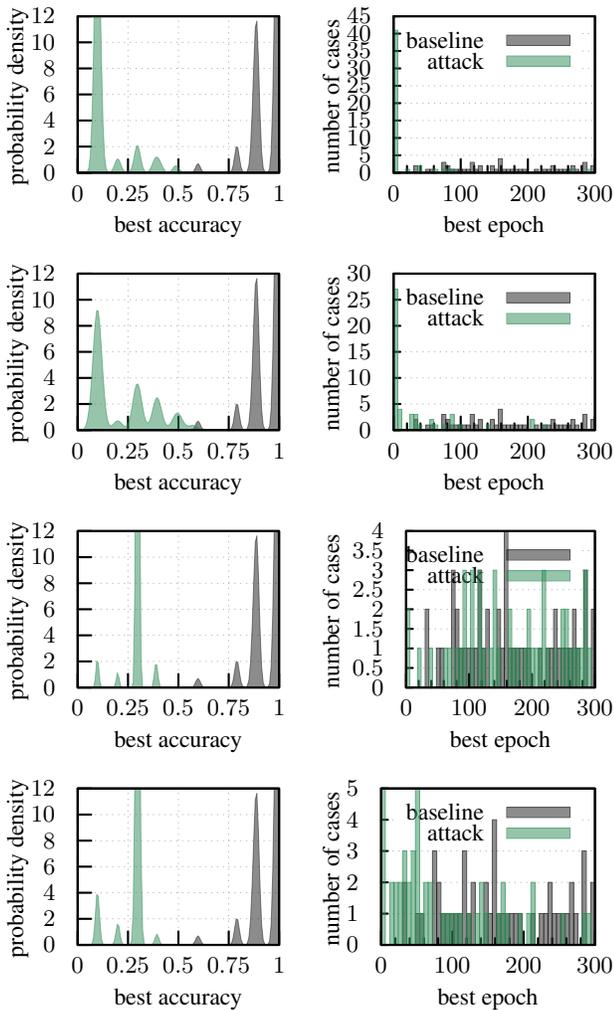

  \input{gp/MNIST_he_Poison_58_lr01}
  \input{gp/MNIST_he_Poison_58_lr005}
  \input{gp/MNIST_he_Poison_58}
  \input{gp/MNIST_he_Poison_58_lr0005}
  \caption{The shift (set to $4$) attack in relation to learning rate on the MNIST data. From top to bottom: learning rate $0.01$, $0.005$, $0.001$ (default) and $0.0005$.}
    \label{fig::shiftLr}
\end{figure}

\section{Full StackOverflow/StackExchange Study}\label{app:study}
We search for ``neural networks low accuracy'', ``neural networks bad performance'', ``neural networks bad accuracy'', and ``neural net fail''
on two popular Q\& A sites for programming issues,  \url{stackexchange.com} and \url{stackoverflow.com}\footnote{Carried out on November 25th 2019 and January 13th 2020, dataset available upon request.}.
Due to fine nuances in meaning, we do not automate the analysis of the 332 posts.
Further due to privacy concerns, we remove all user names from the stored posts and do not carry out any analysis related to users.
We do not count questions without replies (22), duplicates (3), and unrelated questions (185). We consider a question unrelated if it is
\begin{enumerate}
\item a high level question, e.g. \emph{what performs better, neural networks or ensembles}, or \emph{how to deal with missing data},
\item an implementation question, e.g. \emph{In Tensorflow estimator class, what does it mean to train one step?} or \emph{How to use smac for hyper-parameter selection},
\item very specific to an application, e.g. \emph{discussing how to improve the contours of a FedEx logo detector} or \emph{which algorithm to use to block/unblock a gate for vehicles},
\item a question about a specific error message, e.g. \emph{Assertion `cur\_target $>=$ 0 \&\& cur\_target $<$ n\_classes' failed. [...] Any ideas?} or \emph{ValueError: Tensor Tensor("dense\_2/Softmax:0", shape $=$(?, 2 ) , dtype$=$float32) is not an element of this graph.
[...] Any ideas on why this causes this error?}.
\item a question that is of philosophical nature or entirely unrelated to machine learning.
\end{enumerate}

On the remaining relevant questions, we distinguish the \textbf{overfit}ting scenario (high train accuracy, low test accuracy, 21 questions), and a \textbf{bad} performance category (both are low, 115). As some posts are ambiguous (just reporting ``bad accuracy'', 23), we list these separately as \textbf{unclear}.

In each of the above groups, we collect topics mentioned in the posts and categorize them to give a better overview.
We only count one suggestion per category, e.g. if a user writes ``use more data and split the data in a random fashion'', this counts once in the data category.
We then compute the percentage, e.g. for how many percent of the questions this has been suggested.
Hence, 100\% in the data category implies that for each question, people made a suggestion concerning data.
We summarize our results in Table~\ref{table:overflowStudy}.

As expected, for the category overfitting, most replies indicate to use more data ($>$70\%).
Secondly, suggestions concerning the model size prevail.
In the unclear category, most posts concern the data as well, with a tie between changes in the optimizer, momentum etc. and suggestions to fix bugs.
In the bad case, the largest fraction fall as well into the data category, followed tightly by suggestions concerning the architecture of the model.
This also entails changing the initializer---yet we hope to have convinced the reader in particular in this appendix that neither changing the architecture, nor changing the activation, optimizer or initializer will alleviate the attack.

\begin{table*}[t]
  \footnotesize
  \centering
	\caption{Suggestions to fix bad accuracy. Percentage denotes how often reply was given in which setting.}\label{table:overflowStudy}
	\begin{tabular}{@{}l rrrrrr@{}}
	    \toprule[1.5pt]
		Suggestion & Overfit & Bad & Unclear  \\
        \midrule
        More data, sanity check data, data split, re-weight data & 71.4\% & 31.3\% & 61\%  \\
        Train longer & $\emptyset$ & 4.3\% & 13\% & \\
        Change optimizer, loss, batch size, momentum & 14.3\% & 27.8\% & 17.4\%  \\
        Larger/smaller model, different type of layers, change initializer  & 33.3\% & 30.4\% & 26.1\%  \\
        Use regularizer, cross validation, batch normalization, etc.  & 43\% & 14.8\% & 17.4\%  \\
        Use other classifier than DNN & 4.8\% & 7\% &  13\%  \\
        Bugs in logic or implementation &   19\% & 32.2\% & 26.1\% \\
        \addlinespace
        Number of questions evaluated & 21 & 115 & 23  \\
		\bottomrule[1.5pt]
	\end{tabular}
\end{table*}

In total, we find six posts which could potentially lead to the discovery of our adversarial initialization (We also give the rating on the Q\&A which indicates the relevance of the post in the eyes of the community):

\begin{enumerate}
\item \small{\texttt{Check your loss function, weight initialization, and gradient checking to make sure the back-propagation works in an appropriate manner}} (Scenario overfitting, rating: 1)
\item \small{\texttt{your error gradient doesn't reach initial layers! (you can check this by plotting histograms of weights in tensorboard)}} (Scenario bad performance, rating 0)
\item \small{\texttt{Look at individual layers. [...] look for layers which have suspiciously skewed activations (either all 0, or all nonzero)}}(Scenario bad performance, rating 167)
\item \small{\texttt{Did you check if the parameters get updated after optimizer.step()?.}}(Scenario bad performance, rating 0)
\end{enumerate}
Although all or many of these posts potentially give away the attack, the setting described is generally not the one that is caused by our attack.
One concerns a setting where over-fitting takes place.
In our attack, training and test accuracy generally do not diverge as the resulting small model-size prevents memorization.
The other three posts reply to cases where the models do not learn at all, a scenario that our attacker tries to prevent to remain stealthy.

Two replies, however, are posted in the correct setting:
\begin{enumerate}
\item \small{\texttt{This isn't a very good answer (thus why it's a comment) but back when I was studying neural nets you need to double, triple and quadruple check that all your variables and functions are doing and storing what you intend. One single misplaced calculation will mess up the entire system resulting poor results and ripped out hair follicles. Good luck}} (Scenario Unclear, rating 0)
\item \small{\texttt{Gradient check your implementation with a small batch of data and be aware of the pitfalls.}}(Scenario bad performance, rating 1)
\end{enumerate}
Concerning the first reply, it remains an open question what a user would check in detail given these broad instructions.
In the second case, the user points to gradient checking.
This method assumes that the gradient implementation is having a bug---in our setting, however, the bad gradients are a consequence of the small weights, and mathematically correct. Yet, inspection might give away the attack.
We conclude that the victim is more likely to search for more data, or change something about the model, than spot the attack.

\end{document}